# SIMsim: An End-to-End Simulation of The Space Interferometer Mission

David L. Meier*, and William M. Folkner

Jet Propulsion Laboratory, California Institute of Technology

**ABSTRACT**

We present the basic elements and first results of an end-to-end simulation package whose purpose is to test the validity of the Space Interferometer Mission design. The fundamental simulation time step is one millisecond, with substructure at 1/8 ms, and the total duration of the simulation is five years. The end product of a given "wide-angle" astrometry run is an estimated grid star catalog over the entire sky with an accuracy of about 4 micro-arcseconds.

SIMsim is divided into five separate modules that communicate via data pipes. The first generates the 'truth' data on the spacecraft structure and laser metrology. The second module generates uncorrupted fringes for the three SIM interferometers, based on the current spacecraft orientation, target stars' positions, etc. The third module reads out the CCD white light fringe data at specified times, corrupting that and the metrology data with appropriate errors. The data stream out of this module represents the basic data stream on the simulated spacecraft. The fourth module performs fringe-fitting tasks on this data, recovering the total path delay, and the fifth and final module inverts the entire metrology/delay dataset to ultimately determine the instantaneous path delay on a fiducial baseline fixed in space. (Pathlength feed forward is used every few milliseconds to re-position the interferometer to keep the fringes in the delay window.) The average of all such delays over an integration time (typically 30s) is reported as one of several hundred thousand measured stellar delays over the five-year period, which are then inverted to produce the simulated catalog.

Future plans include taking into account more sources of error from the SIM error budget and including narrow angle observations in the observing plan.

## 1. INTRODUCTION

In order for the Space Interferometer Mission to achieve its science goals, it must be able to measure astrometric delays on the science baseline to picometer accuracy over an observation time (typically 30 seconds on a 12[th] magnitude star). During that time the spacecraft will undergo quite a bit of internal and external motion on lengthscales that are orders of magnitude larger than the required delay tolerances. In order that the data can be coherently averaged over an observation time, all such motions must be tracked, and compensated for, using laser metrology and guide star delay measurements on the guide baselines.

Will the system work and obtain the science as advertised? This question really has two parts. The first question concerns the performance of the system components, and has been discussed in several papers earlier in this meeting [1-8]. The second question concerns how well the system will work, given that all the parts work as designed. This paper deals with this latter question. It should be noted that SIMsim is a "medium" fidelity simulation of the mission. That is, it takes into account on-board systems and their intermediate data products, real-time processing, and short time-scale (1 ms) phenomena. In this regard, it has a higher fidelity than, say, the current wide-angle simulations being done by the SIM science planners [9], and therefore can be used to verify key assumptions of the error budget. However, SIMsim does not attempt to simulate the detailed physics on every level. For example, it does not perform diffraction analysis of the individual optical components, nor does it even do ray tracing of the fundamental ray. Instead, it incorporates into its framework simplified representations of results from such "high" fidelity simulations, which are performed by other

*David.L.Meier@jpl.nasa.gov; Jet Propulsion Laboratory, 238-332, 4800 Oak Grove Drive, Pasadena, CA, USA, 91001

members of the engineering team. In this way, SIMsim can address issues of astrometric accuracy and achieve a reasonable degree of fidelity while still addressing the global, 5-year problem. SIMsim should be viewed as only one simulation component in a suite of tools necessary for understanding the Space Interferometer Mission.

The primary goals of SIMsim are to verify that SIM will meet its science goals, both wide-angle and narrow-angle, if the instrument performs as specified. It is not used to design individual components. In addition, it is hoped that SIMsim will be helpful in determining downlink telemetry requirements, in defining which data processing algorithms are implemented onboard and which are performed on the ground, and in investigating different calibration scenarios.

## 2. SIMsim Details: A Review of How SIM Works

The basic astronomical measurement of SIM is the average delay on a "fiducial" science baseline over an integration period. Science delay measurements are made on a timescale long enough to obtain reasonable signal over the readout and photon noises, but short enough so that spacecraft motions do not smear the fringes to the point of being undetectable. The process is then repeated many times over the entire integration period and the results averaged to a single reported delay measurement.

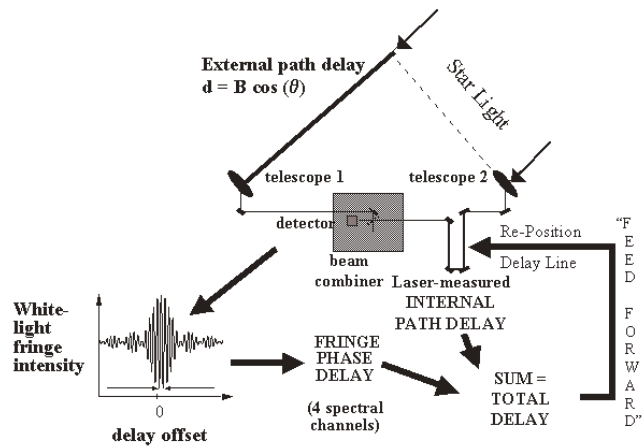

Figure 1. Schematic diagram of an optical interferometer with feed-forward

Figure 1 shows a schematic of a standard optical interferometer. A position of the target star that is not exactly perpendicular to the baseline will generate a geometric delay between the two telescopes of the interferometer. In order to measure this delay, one of the light paths is artificially delayed by internal "delay lines" and then interfered with the other (non-delayed) optical path. In general, the positioning of the delay lines will not be perfect, so there will be some residual delay that can be measured by the interference pattern that is generated. The sum of these two delays (internally-measured and white-light-determined) gives the best approximation to the true external interferometer delay that we wish to measure. For the two SIM guide interferometers, this process should take about one millisecond, assuming that bright, 7[th] magnitude, guide stars are used. For the SIM science interferometer, it will take considerably longer to determine the total path delay, because the target is potentially much fainter (*e.g.*, 12[th] magnitude).

Between each millisecond determination of the guide delays, the spacecraft will undergo a variety of motions (rotation, bending, vibration), which will alter the external path delay. Very quickly, in just a few milliseconds, the principle interferometer fringe can drift out of the delay window; it therefore would be impossible to determine the residual white light delay unless the delay lines were appropriately moved to compensate for the change in external delay due to the spacecraft motion. The use of the guide information to re-position both the guide and science interferometer delay lines is called pathlength "feed-forward". This is essentially a feedback process, but the feedback information is used at a future time (1 millisecond later). The most important result of the feed-forward system is that it maintains the faint fringes on the science baseline so they can be coherently integrated for many milliseconds---long enough for the fringes to be dominated by photon noise rather than readout noise.

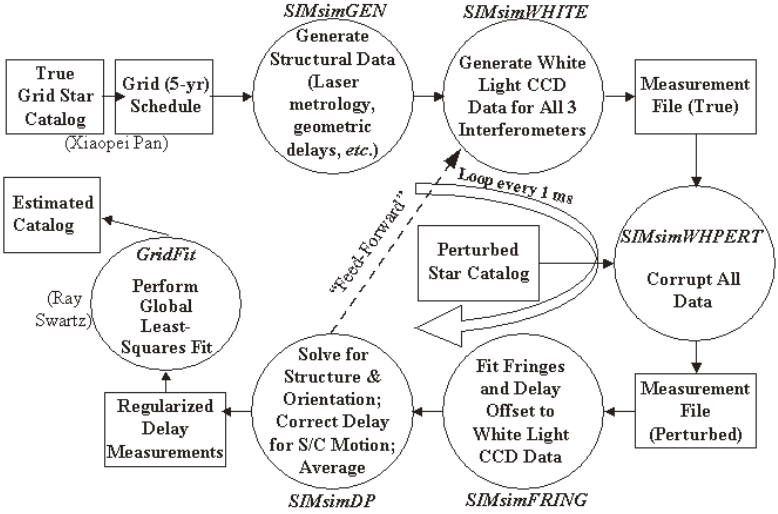

Figure 2. Block diagram of SIMsim modules and data flow

The purpose of SIMsim is to simulate this process in some detail, taking into account spacecraft motions, photon counting and fringe detection, laser metrology errors and biases, calibration errors, and feed-forward issues. The effects of these errors on the global astrometric accuracy then can be evaluated. Figure 2 shows a block diagram of SIMsim, depicting each of the six modules as circular balloons and the data products, intermediate and final, with rectangles. Generally, SIMsim can be viewed as a data pipe, and five of these six programs are run simultaneously in that fashion. Note, however, the feed-forward link, in which information is passed back to the place where the delay lines are positioned and used in the next time step. A full 5-year, 1kHz SIMsim run processes over 400,000 30-second grid star observations. This generates more than $10^{10}$ one-millisecond datasets, each of which is analyzed and used in the feed-forward process. A dataset analysis involves solving six least-squares problems (obtaining white light phases, spacecraft structure from laser metrology, and spacecraft orientation in inertial space). Over 200 TB of data pass through the pipes in Figure 2, and a 5-year mission takes about 2 weeks to simulate on a 14-processor, 400MHz Sun Enterprise 6000.

Below each program step is discussed in some detail.

## 2.1. Step 0: Construct catalogs and schedules

The first step of the simulation is to generate a complete 5-year schedule of observations off-line. This has been done by R. Swartz and X. Pan. Positions of date are needed for star targets, both science and guide, and observation times and durations are generated. Spacecraft and baseline pointing vectors also are needed. Generally the targets are not real stars but rather simulated catalog stars whose positions are known exactly (the "truth" catalog). Two schedule files are then generated---one with the exact star positions of date and one generated in the same manner, but from a perturbed catalog, with position errors of ~0".02 for the guide stars and 0".1 for the ~3000 grid (science) stars. The necessity for both a true and perturbed schedule will become apparent in the discussion of the other modules below. The resulting global 5-year astrometric campaign schedule consists of about 27,000 "tiles", each with about 15 grid star observations of 30 seconds each. These tiles form a linked network around the entire sky, as shown in Figure 3. A full least-squares solution, with 5 parameters per star (ecliptic longitude and latitude, proper motion in each, and parallax) and about

400,000 delay measurements will be performed after SIMsim finishes to generate the estimated catalog. The difference between this and the truth catalog will determine the global wide-angle accuracy of the mission.

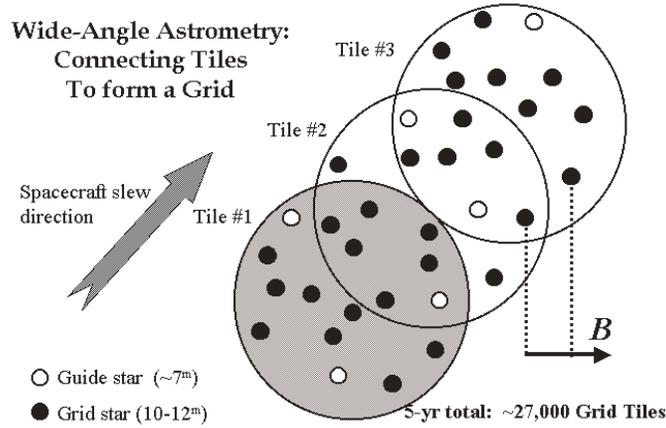

Figure 3. Wide-angle astrometry observing strategy. **B** is the sample baseline orientation. Shaded region shows the current tile being observed.

## 2.2. Step 1: Generate spacecraft structural data (SIMsimGEN)

Steps 1 through 5 describe the simulation processing performed each millisecond for each star delay measurement. The process begins with the computation by program *SIMsimGEN* of the true spacecraft structure, without errors, but with free-space and vibrational motions imposed. The current version of SIMsim still uses the "SIM Classic" design as shown in Figure 4, but we currently are in the process of converting to the new SIM Reference Design, which will have all telescopes near the ends of the spacecraft structure and no external metrology kite. The motions imposed include random vibrations of the laser-reflecting corner cubes, with a period of ~1 ms and an amplitude of ~1 nm. The spacecraft also is rocked in each of the roll, pitch, and yaw axes successively with a period of 20 s and an amplitude of ~1 arcsecond. The purpose of this simple spacecraft model is to produce dramatic changes in the guide (and science) delays, which must be detected by the on-board data processing, and compensated for, so the thousands of measured science delays can be averaged to a common fiducial baseline.

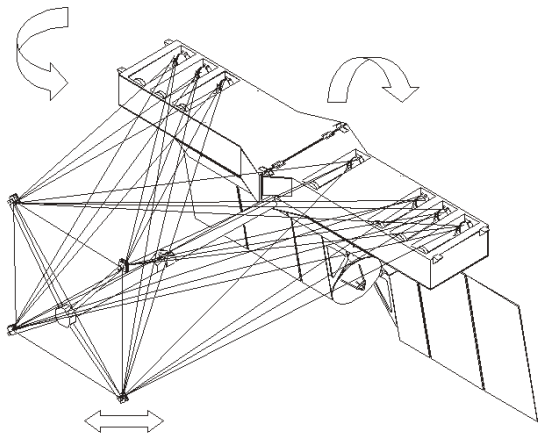

Figure 4. SIM Classic spacecraft structure, showing kite, laser metrology beams, and imposed rocking motions

After generation of the spacecraft model, exact measurements are generated of 1) laser metrology between corner cubes, including kite-to-siderostats and kite-to-kite, and 2) geometric delays ($d_{geometric} = \mathbf{B} \cdot \mathbf{S}$) on the guide and science baselines, where $\mathbf{B}$ is the baseline vector and $\mathbf{S}$ is the star direction unit vector. These are piped to the next program in the sequence.

### 2.3. Step 2: Generate white light CCD data (SIMsim WHITE)

The process of generating the true dataset is completed by program *SIMsimWHITE*, which generates the photon fluxes measured by a CCD as the lengths of the interferometer delay lines are rapidly changed to sample the white light fringes. In order to do so, the length of the interferometer delay line must be set to a value close to, but not necessarily equal to, the interferometer geometric delay. Initially, this delay offset is taken to be a value typical of the errors in the white light fringe estimation (20-30 nm). However, after the first millisecond, the delay line length is set by the feed-forward process, described more fully in section 2.6. It is very important to note that, in general, the delay line length set by the feed forward process will have errors and biases. The errors will result in a persistent delay offset that hopefully will be in the delay window. The unknown delay bias, however, will be very large --- of order several micrometers. In practice, the bias will be compensated for by the process of "fringe acquisition", to make certain the principle interference fringe is in the delay window, and then by maintaining interferometer "lock" throughout the star observation. We simulate this process in the following simple manner. *SIMsimWHITE* assumes that lock has been established after the first millisecond, compares the true delay line length with the fed-forward value, and computes a single constant delay bias $d_{bias}$ to be used for the entire 30 second observation. The bias is then applied to all future delay line settings to determine the true delay line setting to be used by *SIMsimWHITE*

$$d_{line} = d_{line,fed\text{-}forward} - d_{bias} \tag{1}$$

With this procedure, fringes remain within the delay window (subject to white light and other random errors in the system), and the data processing program (*SIMsimDP*) remains unaware of the true lengths of the delay lines.

Once the delay line length has been set, white light fringes are generated by modifying its length with a simulated PZT that strokes 820 nm each millisecond. Eight (8) PZT positions are read out each millisecond, with the zero point centered on the intended delay line length $d_{line}$ ($-410\,\text{nm} < x_{PZT} < 410\,\text{nm}$), and the photons are grouped into four (4) wavelength bins. The resulting fringes are computed from the relation

$$F(\lambda, x_{PZT}) = F_0(\lambda)\{1 + V(\lambda)\cos[2\pi(x_{PZT} + (d_{geometric} - d_{line}))/\lambda]\} \tag{2}$$

where $F$ is the true fringe photon flux, $\lambda$ and $x_{PZT}$ are the photon wavelength and relative PZT offset, $F_0$ is the average flux, and $V$ is the source visibility (as affected by resolution, diffraction, scattering, *etc.*). These photon fluxes are combined with the input metrology measurements to produce the final true measurement file for this millisecond. This data is piped to the next program *SIMsimWHPERT*.

At this point a word needs to be said about star magnitudes and photon fluxes. X. Pan has looked at the photon throughput of the SIM (Classic) optical system and found that a star of about 7[th] magnitude will produce about 960 counts on the CCD every millisecond over the entire SIM bandpass. Dividing this among the $4 \times 8 = 32$ PZT/wavelength bins, and scaling this to magnitude 7, we have the following general expression for the photon flux $F_0$ for any star

$$F_0(\lambda) = 30.0 \times 10^{(m-7)/2.5} \tag{3}$$

where $m$ is the stellar magnitude. For simplicity we have made the assumption that stellar spectral type is the same for all stars --- a reasonable assumption for grid stars, which all will be K-type giants. Further studies of the SIM bandpass by X. Pan have shown that, if the standard photometric red ("*R*") magnitude is used for $m$, then expression (3) is nearly independent of stellar spectral type (although, of course, only at a wavelength $\lambda \sim 700\,\text{nm}$). We therefore consider all magnitudes used in SIMsim to be *R* magnitudes.

## 2.4. Step 3: Corrupt all data with random errors (SIMsimWHPERT)

All errors are added at this step. The true star positions are replaced with ones from the perturbed catalog. Laser metrology measurements are corrupted with random errors (~1.3 nm), fixed biases (~3 μm), and errors that continue to random walk away from their initial zero value (~800 pm in 5 hours, or ~1.7 nm per day). Delay line metrology measurements also are corrupted with similar errors, plus field-dependent calibration errors are added as well. These result because delay line tracks will not be linear at the picometer level over the more than ±1 m of travel necessary to cover the 15 degree field of regard of the science baseline. Specification of this calibration error function is currently an active field of study. We used two simple examples: a random function of position and a systematic, cubic function of angular distance from the center of the field of regard. Errors are also added to the determination of the roll pointing vector (~0.04 arcseconds). The roll pointing direction will be estimated from siderostat and beam compressor gimbals, and will be instrumental in determining the third spacecraft Euler angle (the other two being determined by the guide interferometers). Finally, the white light fringes are truncated to whole photon counts and corrupted with pseudo-Poisson noise ($\sqrt{N}$). Photon readout is also performed in *SIMsimWHPERT*. Guide interferometers are read out every millisecond, while the science interferometer is read out only after 30 or more photons are collected per wavelength/PZT bin. Readout noise is ±3 counts. Figure 5 shows example fringe data in the four chosen wavelength bins. The output of *SIMsimWHPERT* represents the output data stream on the SIM spacecraft.

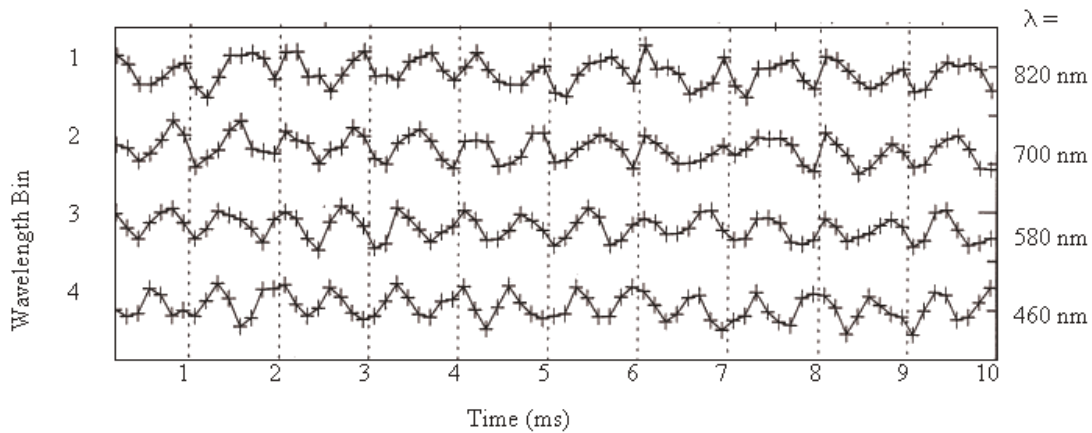

Figure 5. Sample simulated white light fringe data

## 2.5. Step 4: Fit fringes and delay offsets to white light CCD data (SIMsimFRING)

The next two program modules process the simulated SIM data in order to ultimately determine 1) the best position for the delay lines at the next time step and 2) the correction to the measured science delay needed to reference it to a fiducial baseline fixed in space. The first is performed each time the guide interferometers are read out (every millisecond) while the second is done only when the science interferometer is read out (roughly every 100 milliseconds). The first step in each of these is to reduce the white light data from each interferometer to a single estimated fringe delay offset. This offset will be added to the delay line length determined by the laser metrology to compute a total estimated path delay.

The process of estimating the fringe delay is performed by *SIMsimFRING*. A fringe model similar to that employed to generate the fringes (equation 2) is used to filter a least-squares solution for parameters $F_0$, $V$, and the phase offset $\phi$ for the four wavelength bins

$$F(\lambda, x_{PZT}) = F_0(\lambda)\{1 + V(\lambda)\cos[k(\lambda) x_{PZT} + \phi(\lambda)]\} \tag{4}$$

where $k(\lambda) = 2\pi/\lambda$. (In this simulation the PZT positions and effective wavelengths [or $k$'s] of the bins are assumed to be known. In general, however, these will need to be solved for, and monitored. Techniques for fringe estimation with real interferometers and real data are being developed for the SIM Micro-Arcsecond Metrology [MAM] testbed. A more sophisticated fringe model then will be incorporated into SIMsim when the MAM task is completed.) When the photon count in each bin is low, as is the case here, a fringe phase bias is introduced. Milman and Basinger[11] have developed a correction for this bias, which we apply to our phase solutions before computing the fringe delay.

The residual fringe delay $\delta d$ is estimated in *SIMsimFRING* using the phase delay technique, which is considerably more accurate than the group delay technique when the delay offset is less than ¼ wavelength[10]

$$\delta d_{white\_light} = \partial \phi_{fringe} / \partial k = \Sigma_i k_i \phi_i / \Sigma k_i^2 \qquad (5)$$

where the index $i$ runs over the four wavelength bins. In the case of the guide interferometers, this residual delay simply is added to the instantaneous measured delay line length to obtain the total reported delay. In the case of the science baseline, it is added to the laser-metrology-measured delay line length *averaged over the readout time*. The output data stream now has a format identical to that output by *SIMsimGEN*, but with all random, bias, and white light errors now included.

## 2.6. Step 5: Solve for spacecraft structure & orientation, delay feed-forward, and delay regularization (SIMsimDP)

Processing of the metrology data is the most CPU-intensive step and is divided into three parts: 1) computation of the expected future delay on the science baseline, 2) filtering of high-frequency noise from that delay and feeding that result forward to *SIMsimWHITE* to re-position the delay line, and 3) computation of a regularized science delay on a common fiducial baseline, averaged over the ~30 second integration time. All of these are done by *SIMsimDP*.

### 2.6.1. Step 5.1: Compute expected science delay from spacecraft structure

Computing the expected science delay each millisecond involves two sub-steps. First, the spacecraft internal structure (positions of corner cube fiducials) is estimated in the spacecraft reference frame from the laser metrology. For SIM Classic there are 38 laser metrology measurements and 6 geometric constraints on the coordinates. The 36 parameters estimated are the three coordinates of the 12 fiducials (4 kite, 7 siderostats, and one roll fiducial). We find that, despite the μm-level bias, there is enough geometric strength in the laser metrology to obtain a unique solution to this problem, with the metrology triangles closing to within the nanometer measurement error. Once the on-board structure is determined, its orientation in inertial space is determined from the guide interferometers and roll estimator. This problem has four constraints (two guide delays, one roll angle, and the length of the roll estimator, determined by metrology) and three unknowns (the three Euler angles). The resulting instantaneous knowledge of the inertial orientation of the science baseline, plus an *a priori* knowledge of the position of the target science star, then give an estimate of the expected science delay every millisecond

$$d_{expected} = \mathbf{B}_{computed} \bullet \mathbf{S}_{catalog} \qquad (6)$$

Note that this is *not* a measured science delay. It involves neither the science delay line length nor the estimated fringe delay (which is determined only every ~100 ms), but it is the best estimate of the science delay that we have until an actual value can be measured.

### 2.6.2. Step 5.2: Filter expected science delay and feed it forward to re-position the delay line

The above estimated science delays contain a significant amount of random noise, which we filter out using a simple averaging scheme. We find that averaging times of order 10(+6,-2) ms work well, along with a linear predictor. The science delay at the next time step, then, is given by

$$d_{pred}(t) = <d_{expected}>_{t1} + (t-t1)[<d_{expected}>_{t2} - <d_{expected}>_{t1}]/(t2-t1) \quad (7)$$

where $t2-t1 \sim 10$ ms and angle brackets <> denote time averaging. $d_{pred}$ is then piped back to *SIMsimWHITE* every millisecond to re-position the delay line (see Figure 2), and the predictor is re-initialized every ~10 ms. This process is continued for the full ~100 ms science readout time, at the end of which an actual science fringe delay is determined. Figure 6 plots the residual delay $\delta d_{white\_light}$, determined every 100 ms, as a function of time for a typical star observation. The plot shows the effectiveness of the feed-forward system. Despite a variety of spacecraft motions and errors, the delays remain within ~40 nm (about 0.05 wavelength) of the center of the center of the delay window. (The time-dependent structure seen is due to non-linear spacecraft motion that is not handled by the linear predictor in equation 7. Note that the present feed-forward algorithm results in a delay line that is set consistently ~20 nm too short, so a better algorithm therefore may be possible.) The total measured science delay, determined every ~100 ms, is found by adding these residual delays to the length of the delay line, determined by laser metrology.

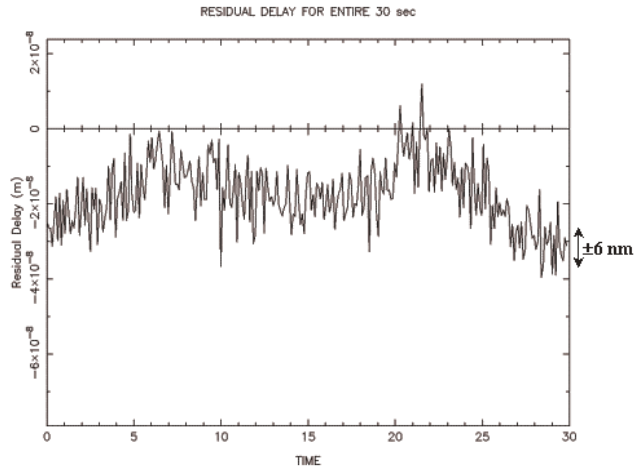

Figure 6. Measured residual science delay *vs.* time for an entire star integration period

### 2.6.3. Step 5.3: Compute regularized science delay and average over the integration period

The total measured delays from Step 5.2 have a considerable amount of structure in them, due mainly to the spacecraft motion. In the current SIMsim simulations, for example, they vary by up to 70 μm over the 30 seconds! However, most of this structure can be eliminated by referencing each measured delay to a common fiducial baseline $\mathbf{B}_{reg}$ that is fixed in inertial space. The choice of the regularizing baseline is somewhat arbitrary, as the global grid solution will re-fit for new baseline parameters and for a constant tile position shift. $\mathbf{B}_{reg}$ can be the average baseline over the integration time, which can be determined only after the fact, or it can be the initial orientation of the baseline. Formally, the error introduced by the regularization process ($\delta\mathbf{B} \cdot \delta\mathbf{S}$) can be no more than about a microarcsecond, which restricts excursions in $\mathbf{B}$ to a few hundred μm for science stars with position errors of order 0".1. Here we have chosen $\mathbf{B}_{reg}$ to be the baseline initially specified in the observing schedule for the current tile. The instantaneous regularized delay is computed using the following equation

$$d_{reg}(t) = d_{measured} + <(\mathbf{B}_{reg} - \mathbf{B}_{computed}) \cdot \mathbf{S}_{catalog}>_{readout}$$

$$= <d_{line}>_{readout} + \delta d_{white\_light} + \mathbf{B}_{reg} \cdot \mathbf{S}_{catalog} - <d_{expected}>_{readout} \quad (8)$$

The final, single regularized delay to be reported for this single star observation is

$$d_{reg} = <d_{reg}(t)>_{integration} \quad (9)$$

averaged now, not over the readout time, but over the entire integration time. Figure 7 shows a plot of the three hundred $d_{reg}(t)$ (minus a constant value) every 100 ms for the same 30 second star observation depicted in Figure 6. The regularized delays are now quite stable, with an RMS of only 6 nm. When averaged, $d_{reg}$ has an uncertainty in the mean of only 6 nm/$\sqrt{300}$ = 350 pm, or about 7 µas on a 10 m baseline. This is well within the SIM wide angle error budget, but it should be noted that we have not yet simulated all the error sources expected in the SIM spacecraft.

This completes the processing of one star observation. The simulation then continues, beginning again with a new target science star and developing a new regularized delay for it. After a tile of about 15 stars is completed, a new tile is begun and completed until all 27,000 tiles have been observed over the 5 year period, with a total runtime of about 2 weeks on a modest parallel enterprise server. This results in about 400,000 stellar delays reported for the 3000 grid stars. These are placed in a "delay measurement file" and passed to the global wide-angle astrometric fitting program.

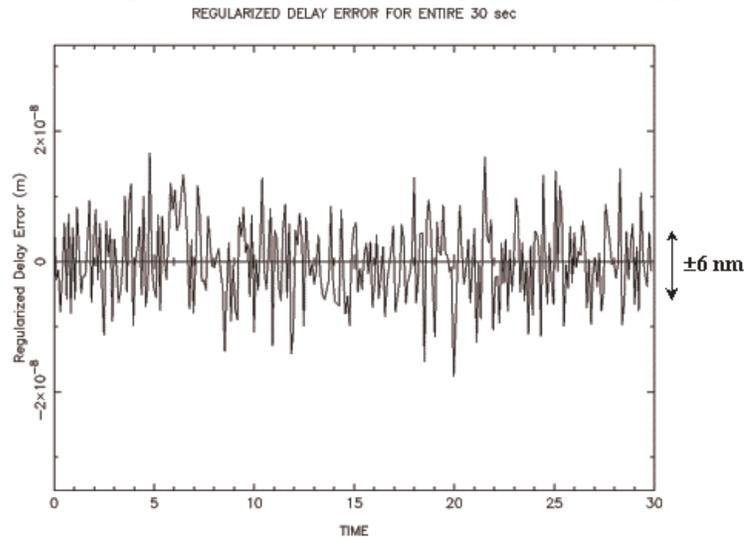

Figure 7. Final regularized delays, minus a constant value

### 2.7. Step 6: Perform a global solution for the estimated catalog (GridFit)

The global astrometric least squares solution is performed off-line from the main SIMsim run, and is discussed by R. Swartz more fully elsewhere [9]. It takes approximately another 8 hours on a 4-processor enterprise server. The 400,000 delays are used to estimate ~15,000 stellar parameters in the catalog and then compared to the original true catalog. For the simulations described here, we find a catalog error of ~1.7 µas in right ascension and declination, with compatible errors in proper motion and parallax, for the case where there is no systematic calibration error in the delay lines.

## 3. Simple SIMsim (SimSIMsim): A Fast 5-year Grid Simulator

While the full version of SIMsim can be used for global grid simulations, a two-week turnaround time for results of a single run is still quite excessive. The 1 kHz version, therefore, has been used primarily for short-term simulations or ones involving considerably fewer than 27,000 tiles, including narrow-angle astrometry, where the 4 µas grid is assumed to already exist. Clearly something faster yet is needed for performing parameter studies of 5-year grid campaigns.

The principal reason for SIMsim's long runtime is the short, 1 millisecond time step, and the principal reason for that is the need to form and analyze white light fringes at every step. If one could approximate the entire white light process with one or two simpler random errors, one could 1) eliminate all the photon and fringe fitting computations (a factor of 2 in speed improvement) and 2) increase the time step by a factor of 10 or 100, and correspondingly reduce all errors applied by the square root of that factor (another factor of 10-100 improvement in speed). After careful analysis of many different star and tile observations, we have been able to do just that. We have found that the main effect of white light and feed-forward processes in SIMsim is to add a single random error to the delay line metrology of

$$\sigma_{white\_light} = 4.8 \times 10^{(m-7)/5.0} (\tau_{data}/1ms)^{-1/2} nm \qquad (10)$$

where $m$ is the star magnitude and $\tau_{data}$ is the desired output time step. For example, with a time step of 10 ms (100 Hz), instead of the full 1 kHz rate, $\sigma_{white\_light}$ = 1.5 nm for a 7th magnitude guide star and 15 nm for a 12th magnitude grid star. We therefore have created a simple version of SIMsim (SimSIMsim), shown in Figure 8, which replaces the white light steps with this single error, replaces *SIMsimWHPERT* with a simpler *SIMsimPERT*, and eliminates the feed-forward loop. This reduces the runtime of a global 15-year simulation to 3 days or less on a single processor, allowing simultaneous multi-parameter grid studies on a parallel machine. Note that only the application of errors is changed in SimSIMsim. The schedule, generation of data, processing of data, and global catalog solution are all the same.

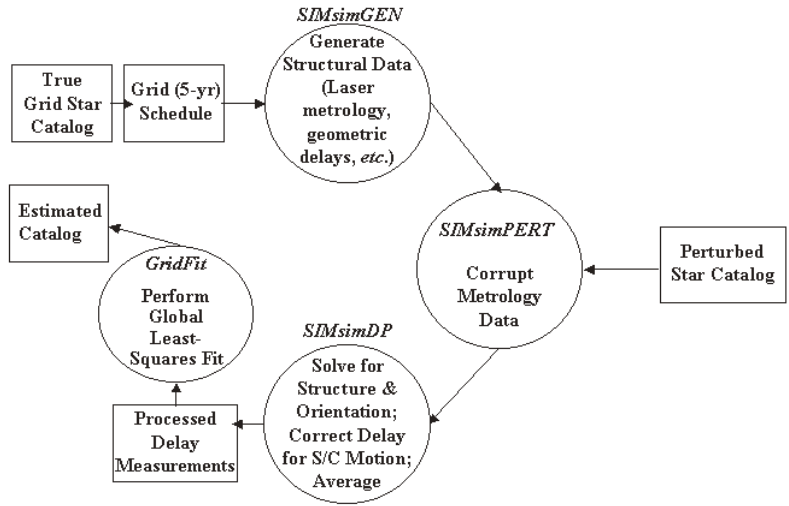

Figure 8. Block diagram of SimSIMsim modules and dataflow

There are, of course, many problems for which the full SIMsim version must be used. These all are ones in which the role of kilohertz effects on the spacecraft are crucial. This includes low-level white light biases (see below), dynamical effects, and narrow-angle observations, which are critically affected by random walk of the laser metrology and other short time scale effects. The other difficulty with SimSIMsim concerns the ~3 μm metrology bias associated with the *a priori* uncertainty in our knowledge of the fiducial corner cube positions. Each inversion of the laser metrology produces a solution for their positions that is accurate to the nanometer level. Over 30 seconds, these 30,000 solutions average down to the five-picometer level. However, if the time step is increased to 10 or 100 ms, the number of metrology solutions averaged is reduced by this factor as well, increasing the effective metrology error on the regularized delay by a factor of 3 to 10. The effects of this can be minimized by using a 10 ms time step and by using the full 1 kHz SIMsim for any narrow-angle studies.

## 4. Sample SIMsim and SimSIMsim Studies

### 4.1. SIMsim studies of biases in the white light phase delay and visibility

When the photon flux $F_0$ is low (<~100 counts per channel per PZT step), the least squares solution for phase $\phi$ and visibility $V$ in equation (4) do not average to the true solution for an infinite number of samples. They are biased. Milman & Basinger[11] and Catanzarite & Milman[12] have developed second-order corrections for these biases that eliminate the effect for reasonable delay offsets and photon fluxes. (For very low photon levels, or very large delay offsets, uncorrected fourth-order effects cause the biases to remain.) We have used SIMsim to test these corrections. This involves observations of a single star, or tile of stars, in which the delay offset is fixed, rather than adjusted by the feed-forward process, white light generated and processed to a derived phase delay offset, and then differenced with the actual offset applied to the delay line to determine a delay bias error. Similarly, the fringe visibilities derived from the least squares solution are differenced with the true visibilities used to generate the original fringes to determine a visibility bias error. Figure 9 shows the results of these two studies. Dashed lines show the uncorrected, biased errors and solid lines show the residual errors after correcting for the bias. The error bars are statistical uncertainties in the large number of samples averaged to produce these curves. The results show that the corrections generate unbiased results within the errors over the range $|\delta d| < 80$ nm both for a $7^{th}$ magnitude guide star, read out every millisecond, and a $12^{th}$ magnitude science (grid) star, read out every 100 milliseconds. The delay offsets that generally occur in SIMsim simulations (see Figure 6) are well within this range, so the Milman-Basinger and Catanzarite-Milman corrections have been incorporated into SIMsim as standard steps in the processing.

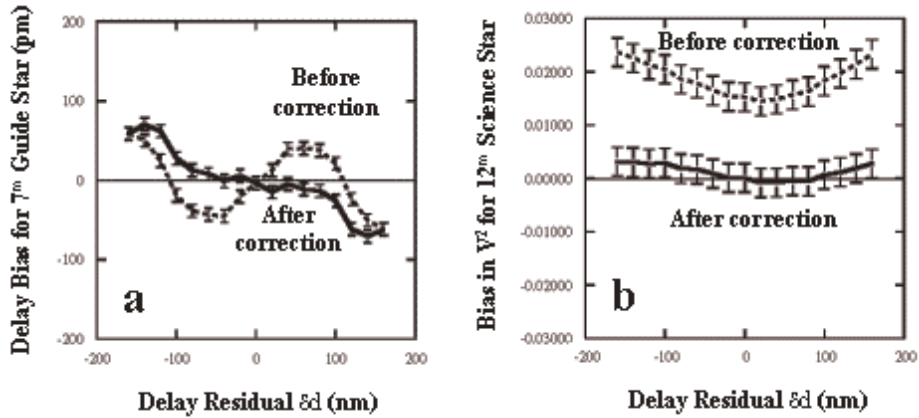

Figure 9. Delay and visibility bias errors before and after the Milman-Basinger and Catanzarite-Milman corrections as a function of the input residual delay

## 4.2. SimSIMsim studies of uncalibrated delay line errors

We have used both SIMsim and SimSIMsim to study the effects of field-of-regard-dependent, uncalibrated delay errors. Two different functional forms of this error have been investigated:

$$\sigma_{cubic} = \sigma_{uncalibrated\_delay}[\delta\phi_{az}^3 + \delta\theta_{el}^3] \qquad (11a)$$

$$\sigma_{random} = \sigma_{uncalibrated\_delay} \mathrm{rand}(\delta\phi_{az}, \delta\theta_{el}) \qquad (11b)$$

where $\delta\phi_{az}$ and $\delta\theta_{el}$ are offsets in siderostat azimuth and elevation from the center of the field of regard (see, e.g., Figure 3) and $\sigma_{uncalibrated\_delay}$ is a constant representing the level of error. Figure 10 shows the derived astrometric error of the global catalog of 3000 stars for standard metrology and white light errors and for different uncalibrated delay errors (0, 0.15, 1.5, and 15 nm). Most of the runs used SimSIMsim, with one of the 1.5 nm runs using the full SIMsim as a check. The results show that 1) large values of the uncalibrated error can dominate, 2) the basic level of the error is much more important than its functional form, and 3) the final catalog error can be approximated by

$$\sigma_{catalog} = 1.7(1 + [\sigma_{uncalibrated\_delay}/450\mathrm{pm}]^2)^{1/2} \mu\mathrm{as} \qquad (12)$$

*if* these runs account for all of the errors in the SIM global astrometric mission. According to equation (12), in order to meet the SIM goal of 4 µas global astrometry, the total uncalibrated error (field-dependent and otherwise) can be no more than 960 pm. Current estimates of $\sigma_{uncalibrated\_delay}$ are of order 150-200 pm, leaving room in the error budget for other terms not yet taken into account by SIMsim.

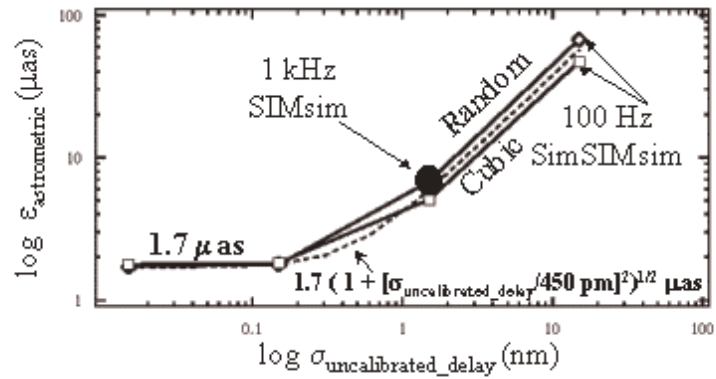

Figure 10. Global astrometric catalog error vs. uncalibrated delay error for different functional forms. The dotted line is a simple "root-sum-squared" approximation to the results.

## 5. Conclusions and Future Studies and Enhancements

SIMsim has been shown to be an effective tool in verifying the capabilities of the Space Interferometer Mission. However, many improvements will be needed, and many more studies performed, before the task can be declared complete. The most important modification will be to convert SIMsim from the Classic design to the current Reference Design. This will be necessary to perform perhaps SIMsim's most important study, narrow-angle astrometry. Narrow-angle science will be affected by detailed calibration errors, random metrology walk, and the details of the Reference Design, including placement of the siderostats, the roll estimator, and the guide star choices. Additional improvements to SIMsim will include better spacecraft modeling, better PZT modeling, adding more important error budget terms, further improvements in speed, and operation on a true supercomputer in the multi-gigaflop range. Future studies will include determining downlink telemetry bandwidth requirements, defining which algorithms will be used for on-board data processing and which can be executed on the ground, and better understanding of field-dependent calibration issues.

## ACKNOWLEDGMENTS

The authors are especially grateful to X. Pan and R. Swartz who performed the catalog/schedule preparation and grid solutions. This research was carried out at the Jet Propulsion Laboratory, California Institute of Technology, under contract to the National Aeronautics and Space Administration.